\documentclass{article}

\usepackage{hyperref}

\usepackage{amsmath, amsthm}
\usepackage[T1]{fontenc}
\usepackage{stix}
\usepackage{libertine}
\usepackage{microtype}

\usepackage{aliascnt}
\newcommand\mynewtheorem[2]{%
  \newaliascnt{#1}{theorem}%
  \newtheorem{#1}[#1]{#2}%
  \aliascntresetthe{#1}%
  \expandafter\def\csname#1autorefname\endcsname{#2}%
}

\theoremstyle{plain}
\newtheorem{theorem}{Theorem}[section]

\mynewtheorem{lemma}{Lemma}
\mynewtheorem{corollary}{Corollary}
\mynewtheorem{proposition}{Proposition}

\theoremstyle{definition}
\mynewtheorem{example}{Example}
\mynewtheorem{definition}{Definition}
\mynewtheorem{convention}{Convention}
\mynewtheorem{remark}{Remark}
\mynewtheorem{notation}{Notation}

\newenvironment{ottdefnblock}[3][]{ \framebox{\mbox{#2}} \quad #3 \\[0pt]}{}

\newcommand{\ottnt}[1]{\mathit{#1}}
\newcommand{\ottmv}[1]{\mathit{#1}}
\newcommand{\ottkw}[1]{\mathbf{#1}}
\newcommand{\ottsym}[1]{#1}

\renewcommand{\ottkw}[1]{\mathrm{#1} }
\renewcommand{\ottnt}[1]{#1}
\renewcommand{\ottmv}[1]{#1}
\renewcommand{\ottsym}[1]{#1}

\usepackage{mysyntax}
\usepackage{myrules}

\newcommand\secref[1]{\autoref{sec:#1}}
\newcommand\figref[1]{\autoref{fig:#1}}
\newcommand\propref[1]{\autoref{prop:#1}}

\newcommand{\LP}{\ensuremath{\lambda^\parallel}}
\newcommand{\LC}{\ensuremath{\lambda^{\parallel\omega}}}
\newcommand{\LHB}{\ensuremath{\lambda^{H\parallel^\Phi}}}
\newcommand{\OSTEP}{\mathrel{\overset{ \varnothing }{ \longrightarrow }{}^\ast}}
\providecommand{\msansA}{\mathsf{A}}
\providecommand{\BbbL}{\mathbb{L}}
\providecommand{\mathampersand}{\&}

\title{Compilation of Coordinated Choice}
\author{Yuki Nishida \and Atsushi Igarashi}

\begin{document}

\maketitle

\begin{abstract}
  Recently, we have proposed \emph{coordinated choices}, which are
  nondeterministic choices equipped with names.  The main characteristic of
  coordinated choices is that they synchronize nondeterministic decision among
  choices of the same name.

  The motivation of the synchronization mechanism is to solve a theoretical
  problem.  So, as a practical programming language, we still want to use
  coordinated choices like standard ones.  In other words, we want to avoid
  synchronization.  Now, there are two problems: (i) practically, it is a bit
  complicated work to write a program using coordinated choices in which
  execution synchronization never happens; and (ii) theoretically, it is unknown
  whether any programs using standard choices can be written by using only
  coordinated ones.

  In this paper, we define two simply typed lambda calculi called \LP{} equipped
  with standard choices and \LC{} equipped with coordinated choices, and give
  compilation rules from the former into the latter.  The challenge is to show
  the correctness of the compilation because behavioral correspondence between
  expressions before and after compiling cannot be defined directly by the
  compilation rules.  For the challenge, we give an effect system for \LC{} that
  characterizes expressions in which execution synchronization never happens.
  Then, we show that all compiled expressions can be typed by the effect system.
  As a result, we can easily show the correctness because the main concern of
  the correctness is whether synchronization happens or not.
\end{abstract}

\section{Introduction}\label{sec:cochoice/introduction}

\emph{Nondeterministic choice}, written \(\ottsym{(}  \ottnt{M_{{\mathrm{1}}}}  \parallel  \ottnt{M_{{\mathrm{2}}}}  \ottsym{)}\) in this paper, is a
rather well-known construct formalizing nondeterministic computation.
Nondeterminism is typically accomplished by the following two operational rules:
\(\ottsym{(}  \ottnt{M_{{\mathrm{1}}}}  \parallel  \ottnt{M_{{\mathrm{2}}}}  \ottsym{)}  \longrightarrow  \ottnt{M_{{\mathrm{1}}}}\) and \(\ottsym{(}  \ottnt{M_{{\mathrm{1}}}}  \parallel  \ottnt{M_{{\mathrm{2}}}}  \ottsym{)}  \longrightarrow  \ottnt{M_{{\mathrm{2}}}}\).  So, for example,
\(\ottsym{(}  1  \parallel  2  \ottsym{)}  \ottsym{+}  \ottsym{(}  3  \parallel  4  \ottsym{)}\) is evaluated into one of the three possible values:
\(4, 5, 6\).

Recently, we had proposed \emph{coordinated choices}~\cite{NishidaI2018},
written \( \ottsym{(} \ottnt{M_{{\mathrm{1}}}} \parallel ^{ \Phi } \ottnt{M_{{\mathrm{2}}}} \ottsym{)} \).  It is a kind of nondeterministic choice but
equipped with a name \(\Phi\).  The characteristic point of coordinated choices
is that nondeterministic evaluation is synchronized among choices of the same
name.  That means, for instance, \( \ottsym{(} 1 \parallel ^{  \msansA  } 2 \ottsym{)}   \ottsym{+}   \ottsym{(} 3 \parallel ^{  \msansA  } 4 \ottsym{)} \) only results
\(4, 6\).  The essential purpose of developing coordinated choices is to solve a
theoretical problem that happens when we try to introduce nondeterministic
choices into a dependently typed system.  In other words, the proposed system
does not have nondeterministic choices.

Now, it is a natural and important demand for practically and theoretically, how
we use coordinated choices like standard ones.  In this paper, we consider how
to simulate usual choices with coordinated choices.  Concretely, we give a
compilation algorithm from a simply typed lambda calculus with nondeterministic
choices into one with coordinated choices; and show an evaluation in the former
calculus is simulated by the latter calculus, and vice versa.


\section{Compilation}\label{sec:cochoice/compilation}

In this section, we introduce the compilation rules and discuss the correctness
of the given rules.  First of all, we recall how a coordinated choice is
synchronized.  The following is an example of a reduction sequence in which
synchronization happens.
\begin{align*}
   \mathtt{add}  \,  \ottsym{(} 1 \parallel ^{  \msansA  } 2 \ottsym{)}  \,  \ottsym{(} 3 \parallel ^{  \msansA  } 4 \ottsym{)} 
  &\OSTEP{}  \ottsym{(}  \mathtt{add}  \, 1 \parallel ^{  \msansA  }  \mathtt{add}  \, 2 \ottsym{)}  \,  \ottsym{(} 3 \parallel ^{  \msansA  } 4 \ottsym{)} \\
  &\OSTEP{}  \ottsym{(}  \mathtt{add}  \, 1 \,  \ottsym{(} 3 \parallel ^{  \msansA  } 4 \ottsym{)}  \parallel ^{  \msansA  }  \mathtt{add}  \, 2 \,  \ottsym{(} 3 \parallel ^{  \msansA  } 4 \ottsym{)}  \ottsym{)} \\
  &\OSTEP{}  \ottsym{(}  \mathtt{add}  \, 1 \, 3 \parallel ^{  \msansA  }  \mathtt{add}  \, 2 \, 4 \ottsym{)} \\
  &\OSTEP{}  \ottsym{(} 4 \parallel ^{  \msansA  } 6 \ottsym{)} 
\end{align*}
In the previous work, we had made two design decision to formalize
nondeterministic choices.  One is that we express the possibility of execution
as of the form of a choice rather than branches of reduction sequences as we
have seen in \secref{cochoice/introduction}.  So, the result of execution above
becomes the choice, which expresses the possible results are only 4 and 6.
Another is that we adopt \emph{call-time} semantics~\cite{HennessyA77}, which
makes nondeterministic choice happening before a function call.  The
distribution of the first argument for \texttt{add} in the first line, for
example, accomplishes this semantics.

The devices that enable synchronization are following rules, which are used in
the evaluation from the second line to the third line.
\begin{center}
\begin{rules}[axiomB]
  \infun
  {\( \ottnt{M_{{\mathrm{1}}}}  \mathrel{\overset{ \Delta  \cup  \ottsym{\{}   \Phi _{+}   \ottsym{\}} }{ \longrightarrow } }  \ottnt{M'_{{\mathrm{1}}}} \)}
  {\(  \ottsym{(} \ottnt{M_{{\mathrm{1}}}} \parallel ^{ \Phi } \ottnt{M_{{\mathrm{2}}}} \ottsym{)}   \mathrel{\overset{ \Delta }{ \longrightarrow } }   \ottsym{(} \ottnt{M'_{{\mathrm{1}}}} \parallel ^{ \Phi } \ottnt{M_{{\mathrm{2}}}} \ottsym{)}  \)}
  \infun
  {\vphantom{\( \ottnt{M_{{\mathrm{1}}}}  \mathrel{\overset{ \Delta  \cup  \ottsym{\{}   \Phi _{+}   \ottsym{\}} }{ \longrightarrow } }  \ottnt{M'_{{\mathrm{1}}}} \)}}
  {\(  \ottsym{(} \ottnt{M_{{\mathrm{1}}}} \parallel ^{ \Phi } \ottnt{M_{{\mathrm{2}}}} \ottsym{)}   \mathrel{\overset{ \Delta  \cup  \ottsym{\{}   \Phi _{+}   \ottsym{\}} }{ \longrightarrow } }  \ottnt{M_{{\mathrm{1}}}} \)}
\end{rules}
\end{center}
So, concentrating on one step of an evaluation, we can easily find that a sufficient
condition to prevent synchronization is that every name of choices in an
expression before the evaluation is distinct.  However, this condition is easily
broken by evaluation since some reduction rules, e.g.,
\( \ottsym{(}  \lambda  \ottmv{x}  \ottsym{.}  \ottnt{M}  \ottsym{)} \, \ottnt{V}  \mathrel{\overset{ \Delta }{ \longrightarrow } }   \ottnt{M}  [  \ottmv{x}  \coloneq  \ottnt{V}  ]   \), duplicate an expression.  In other words, it is
easy to prevent synchronization in the first step of an evaluation, but it is
hard to prevent synchronization in every step after the first step of an
evaluation.

The following reduction sequence shows synchronization caused by duplication of
an expression.  In the first line, recursive call of \(\ottmv{f}\) duplicates the
name \(\ottnt{A}\).
\begin{align*}
  \ottsym{(}   \mu \ottmv{f} . \lambda  \ottmv{x}  \ottsym{.}   \ottsym{(} \ottmv{f} \, \ottmv{x} \parallel ^{  \msansA  } \ottmv{f} \, \ottmv{x} \ottsym{)}    \ottsym{)} \, 0
  &\OSTEP{}  \ottsym{(} \ottmv{f'} \, 0 \parallel ^{  \msansA  } \ottmv{f'} \, 0 \ottsym{)} \\
  &\OSTEP{}  \ottsym{(}  \ottsym{(} \ottmv{f'} \, 0 \parallel ^{  \msansA  } \ottmv{f'} \, 0 \ottsym{)}  \parallel ^{  \msansA  }  \ottsym{(} \ottmv{f'} \, 0 \parallel ^{  \msansA  } \ottmv{f'} \, 0 \ottsym{)}  \ottsym{)} \\
  &\OSTEP{}  \ottsym{(} \ottmv{f'} \, 0 \parallel ^{  \msansA  } \ottmv{f'} \, 0 \ottsym{)}  \OSTEP{} \dots,\\
    \text{where \( \ottmv{f'} \ottsym{=}  \mu \ottmv{f} . \lambda  \ottmv{x}  \ottsym{.}   \ottsym{(} \ottmv{f} \, \ottmv{x} \parallel ^{  \msansA  } \ottmv{f} \, \ottmv{x} \ottsym{)}   \).}
\end{align*}

To avoid the synchronization, the proposed system also has name abstraction
mechanism, i.e., name abstractions \( \Lambda  \alpha . \ottnt{M} \); name applications \(\ottnt{M} \, \Phi\);
and name concatenations \( \Phi_{{\mathrm{1}}} \Phi_{{\mathrm{2}}} \).
Fortunately, thanks to call-by-value semantics, names that will be duplicated
and become a cause of synchronization only occur in the body of lambda
abstractions.  So we abstract names (giving a name containing variables)
occurring in the body of lambda abstractions and instantiate the names as
distinct ones where the lambda abstractions are called.  The following two
evaluation sequences show the effect of this idea.
\begin{align*}
  &\ottsym{(}   \mu \ottmv{f} . \lambda  \ottmv{x}  \ottsym{.}   \Lambda  \alpha .  \ottsym{(} \ottmv{f} \, \ottmv{x} \,  \alpha \mdwhtcircle  \parallel ^{ \alpha } \ottmv{f} \, \ottmv{x} \,  \alpha \mdblkcircle  \ottsym{)}     \ottsym{)} \, 0 \, \mdwhtcircle\\
  \OSTEP{} & \ottsym{(} \ottmv{f'} \, 0 \,  \mdwhtcircle \mdwhtcircle  \parallel ^{ \mdwhtcircle } \ottmv{f'} \,  \mdwhtcircle \mdblkcircle  \ottsym{)} \\
  \OSTEP{} & \ottsym{(}  \ottsym{(} \ottmv{f'} \, 0 \,   \mdwhtcircle \mdwhtcircle  \mdwhtcircle  \parallel ^{  \mdwhtcircle \mdwhtcircle  } \ottmv{f'} \, 0 \,   \mdwhtcircle \mdwhtcircle  \mdblkcircle  \ottsym{)}  \parallel ^{ \mdwhtcircle }  \ottsym{(} \ottmv{f'} \, 0 \,   \mdwhtcircle \mdblkcircle  \mdwhtcircle  \parallel ^{  \mdwhtcircle \mdblkcircle  } \ottmv{f'} \, 0 \,   \mdwhtcircle \mdblkcircle  \mdblkcircle  \ottsym{)}  \ottsym{)} ,\\
  &\text{where \( \ottmv{f'} \ottsym{=}  \mu \ottmv{f} . \lambda  \ottmv{x}  \ottsym{.}   \Lambda  \alpha .  \ottsym{(} \ottmv{f} \, \ottmv{x} \,  \alpha \mdwhtcircle  \parallel ^{ \alpha } \ottmv{f} \, \ottmv{x} \,  \alpha \mdblkcircle  \ottsym{)}    \).}
\end{align*}
In the first sequence, the duplicated name \(\Phi\) causes synchronization.  In
the second sequence, on the other hand, the name of a coordinated choice becomes
different one at every time that a lambda abstraction is applied (so
synchronization does not happen).

\begin{figure}
  \centering
  \begin{align*}
      \lBrack \ottmv{x} \rBrack^{ \alpha }_{ \bar\Phi }   &=  \ottmv{x} \\
      \lBrack \ottnt{e_{{\mathrm{1}}}} \, \ottnt{e_{{\mathrm{2}}}} \rBrack^{ \alpha }_{ \bar\Phi }   &=   \lBrack \ottnt{e_{{\mathrm{1}}}} \rBrack^{ \alpha }_{   \bar\Phi \mdwhtcircle  \mdwhtcircle  }  \,  \lBrack \ottnt{e_{{\mathrm{2}}}} \rBrack^{ \alpha }_{   \bar\Phi \mdwhtcircle  \mdblkcircle  }  \,   \alpha \bar\Phi  \mdblkcircle  \\
      \lBrack \lambda  \ottmv{x}  \ottsym{.}  \ottnt{e} \rBrack^{ \alpha }_{ \bar\Phi }   &=  \lambda  \ottmv{x}  \ottsym{.}   \Lambda  \beta .  \lBrack \ottnt{e} \rBrack^{ \beta }_{ \bar\Phi }   \\
      \lBrack  \mu \ottmv{f} . \ottnt{e}  \rBrack^{ \alpha }_{ \bar\Phi }   &=   \mu \ottmv{f} .  \lBrack \ottnt{e} \rBrack^{ \alpha }_{ \bar\Phi }   \\
      \lBrack \ottsym{(}  \ottnt{e_{{\mathrm{1}}}}  \parallel  \ottnt{e_{{\mathrm{2}}}}  \ottsym{)} \rBrack^{ \alpha }_{ \bar\Phi }   &=   \ottsym{(}  \lBrack \ottnt{e_{{\mathrm{1}}}} \rBrack^{ \alpha }_{  \bar\Phi \mdwhtcircle  }  \parallel ^{   \alpha \bar\Phi  \mdblkcircle  }  \lBrack \ottnt{e_{{\mathrm{2}}}} \rBrack^{ \alpha }_{  \bar\Phi \mdwhtcircle  }  \ottsym{)}  
  \end{align*}
  \caption{Compilation of expressions}
  \label{fig:cochoice/compile}
\end{figure}

Summing up the discussion, concrete compilation rules become the ones in
\figref{cochoice/compile}.  The function \(\lBrack \cdot \rBrack^\alpha_\omega\)
gives the compilation, where \(\alpha\) and \(\omega\) are additional
parameters, called \emph{seeds}, used for generating distinct names during a
compilation.  Note that the same seeds are given for the sub-expressions'
compilation of a choice.  It does not cause a problem because the left and right
sides of a choice are completely separated, i.e., the sub-expressions never
collaborate.

\subsection{Correctness of compilation}
The main contribution of this paper is that we have formally shown that the
compilation is \emph{correct}.  Informally, correctness of a compilation is
stated as---a compiled expression behaves the same as the original expression
does.  This behavioral correspondence between compiled expressions and original
expressions is formally defined as a binary relation, called
\emph{bisimulation}~\cite{Milner1989}, between them.  For instance, if we use the
compilation rules as the relation, what we need to show is stated as the
following two propositions.

\begin{proposition}\label{prop:cochoice/disim1}
  If\/ \(  \lBrack \ottnt{e} \rBrack^{ \alpha }_{ \bar\Phi }  \ottsym{=} \ottnt{M} \) and \(\ottnt{e}  \longrightarrow  \ottnt{e'}\), then \( \ottnt{M}  \mathrel{\overset{ \Delta }{\longrightarrow}{}^\ast}  \ottnt{M'} \) and
  \(  \lBrack \ottnt{e'} \rBrack^{ \alpha }_{ \bar\Phi }  \ottsym{=} \ottnt{M'} \).
\end{proposition}

\begin{proposition}\label{prop:cochoice/disim1}
  If\/ \(  \lBrack \ottnt{e} \rBrack^{ \alpha }_{ \bar\Phi }  \ottsym{=} \ottnt{M} \) and \( \ottnt{M}  \mathrel{\overset{ \Delta }{ \longrightarrow } }  \ottnt{M'} \), then \(\ottnt{e}  \longrightarrow^\ast  \ottnt{e'}\) and
  \(  \lBrack \ottnt{e'} \rBrack^{ \alpha }_{ \bar\Phi }  \ottsym{=} \ottnt{M'} \).
\end{proposition}

However, we can easily find a counter-example for the reduction
\(\ottsym{(}  \lambda  \ottmv{x}  \ottsym{.}  \ottmv{x}  \ottsym{)} \, \lambda  \ottmv{x}  \ottsym{.}  \ottmv{x}  \longrightarrow  \lambda  \ottmv{x}  \ottsym{.}  \ottmv{x}\) as follows.
\begin{align*}
    \lBrack \ottsym{(}  \lambda  \ottmv{x}  \ottsym{.}  \ottmv{x}  \ottsym{)} \, \lambda  \ottmv{x}  \ottsym{.}  \ottmv{x} \rBrack^{ \alpha }_{ \bar\Phi }   &=   \lBrack \lambda  \ottmv{x}  \ottsym{.}  \ottmv{x} \rBrack^{ \alpha }_{   \bar\Phi \mdwhtcircle  \mdwhtcircle  }  \,  \lBrack \lambda  \ottmv{x}  \ottsym{.}  \ottmv{x} \rBrack^{ \alpha }_{   \bar\Phi \mdwhtcircle  \mdblkcircle  }  \,   \alpha \bar\Phi  \mdblkcircle  \\
                            &= \ottsym{(}  \lambda  \ottmv{x}  \ottsym{.}   \Lambda  \beta . \ottmv{x}   \ottsym{)} \,  \lBrack \lambda  \ottmv{x}  \ottsym{.}  \ottmv{x} \rBrack^{ \alpha }_{   \bar\Phi \mdwhtcircle  \mdblkcircle  }  \,   \alpha \bar\Phi  \mdblkcircle 
                              \overset{\Delta}{\longrightarrow}  \lBrack \lambda  \ottmv{x}  \ottsym{.}  \ottmv{x} \rBrack^{ \alpha }_{   \bar\Phi \mdwhtcircle  \mdblkcircle  } 
\end{align*}
A seed is alternated by the reduction.  Unfortunately, weakening the proposition
as follows does not help us, because, if a sub-expression (for example, the
argument part of an application) is evaluated, only the seed for the
sub-expression part is alternated and the whole expression after the evaluation
does not follow the compilation rule.

\begin{proposition}\label{prop:cochoice/disim2}
  If\/ \(  \lBrack \ottnt{e} \rBrack^{ \alpha }_{ \bar\Phi }  \ottsym{=} \ottnt{M} \) and \(\ottnt{e}  \longrightarrow  \ottnt{e'}\), then \( \ottnt{M}  \mathrel{\overset{ \Delta }{ \longrightarrow } }  \ottnt{M'} \) and
  \(  \lBrack \ottnt{e'} \rBrack^{ \alpha }_{ \bar\Phi' }  \ottsym{=} \ottnt{M'} \) for some \(\bar\Phi'\).
\end{proposition}

Ultimately, we believe that the relation cannot be obtained directly by a
compilation algorithm.  That means it is not because of how the compilation
rules are defined.  To prevent coordination of coordinated choice, we need to
give distinct names for each coordinated choice, which is a sub-expression, but
as we have seen in the counter-example, a sub-expression could pop-up to the top
level, which leads us to an inconsistency.

Summarizing the discussion, we could see that coordination does not happen in
the first step of the evaluation for a complied expression; but we cannot
guarantee the property after the first step because an expression can have no
relation to the compilation rules.  So, we take another indirect strategy.
Firstly, we define a relation characterizing expressions that do not cause
coordination and is preserved by an evaluation, and then we show the image of the
compilation is included in the relation.  After accomplishing this, we can
easily obtain a bisimulation relation since the compilation just inserts name
abstractions and applications---an evaluation after compilation just involves
\ruleref{R-Sigma} reduction in some points.


\section{Formal System}\label{sec:cochoice/system}

We formalize the idea as \LP{}, a simply typed lambda calculus with a fix-point
operator and \emph{non-collapse} choices, and \LC{}, a simply typed lambda
calculus with a fix-point operator and coordinated choice; and give a
compilation rule from the former to the latter.

\subsection{Source Language: \LP{}}

\begin{figure}
  \centering
  \input{figures/ssyntax}
  \caption{Syntax of \LP{}}
  \label{fig:cochoice/ssyntax}
\end{figure}

\begin{figure}
  \centering
  \input{figures/ssemantics}
  \caption{Operational semantics of \LP{}}
  \label{fig:cochoice/ssemantics}
\end{figure}

\begin{figure}
  \centering
  \input{figures/styping}
  \caption{Typing rules for \LP{}}
  \label{fig:cochoice/styping}
\end{figure}

The syntax, semantics, and type system of \LC{} are defined as
\figref{cochoice/ssyntax}, \figref{cochoice/ssemantics}, and
\figref{cochoice/styping}, respectively.  Those are standard ones for a simply
typed lambda calculus and with straightforward extensions for a fix-point
operator \( \mu \ottmv{f} . \ottnt{e} \) and non-collapse choices \(\ottsym{(}  \ottnt{e_{{\mathrm{1}}}}  \parallel  \ottnt{e_{{\mathrm{2}}}}  \ottsym{)}\).
``Non-collapse'' means that a choice dose not really choose an alternative (like
\(\ottsym{(}  \ottnt{e_{{\mathrm{1}}}}  \parallel  \ottnt{e_{{\mathrm{2}}}}  \ottsym{)}  \longrightarrow  \ottnt{e_{{\mathrm{1}}}}\)) as coordinated choice does not, but also it never
coordinate alternatives.

\subsection{Target Language: \LC{}}

\begin{figure}
  \centering
  \input{figures/tsyntax}
  \caption{Syntax of \LC{}}
  \label{fig:cochoice/tsyntax}
\end{figure}

\begin{figure}
  \centering
  \input{figures/tsemantics}
  \caption{Operational semantics of \LC{}}
  \label{fig:cochoice/tsemantics}
\end{figure}

The syntax and semantics of \LC{} are shown in \figref{cochoice/tsyntax} and
\figref{cochoice/tsemantics}, respectively.  As far as the semantics, \LC{} is
just a simply typed lambda calculus with coordinated choice, c.f.\ \LHB{}.  An
interesting part of \LC{} is an \emph{effect} system~\cite{NielsonN1999}, which is a kind of type
system.  The effect system of \LC{} estimates names happening during execution
and rejects an expression that will cause synchronization, and so this is the
relation that we have mentioned at the last of \secref{cochoice/compilation}.

We range over effects with a meta-variable \(\varphi\).  Effects of \LC{} are
denoted by regular expressions: the empty set \( \varnothing \); literal strings
\(\Phi\) including the empty string \( \epsilon \); concatenations \( \varphi_{{\mathrm{1}}}   \varphi_{{\mathrm{2}}} \);
alternations \(\varphi_{{\mathrm{1}}}  \ottsym{+}  \varphi_{{\mathrm{2}}}\); and Kleene star \( \varphi ^\ast \).

Function types and forall types are annotated by effects, which express the
effect happening when a function is used.  Forall types \( \forall \alpha .^{ \varphi } \tau \) bind
\(\alpha\) in \(\varphi\) and \(\tau\).

\begin{definition}
  A \emph{closed} term is one in which both free variables and free name
  variables are empty.  We denote closed terms with over-barred meta-variables,
  e.g., \(\bar\Phi, \bar\varphi\).
\end{definition}

\begin{definition}
  We denote the language, a set of names, represented by an effect \(\varphi\) as
  \( \BbbL ( \varphi ) \).
\end{definition}

\subsection{Effect System}

\begin{figure}
  \centering
  \input{figures/ttyping1}
  \caption{Effect system of \LC{} (1): subtyping rules}
  \label{fig:ttyping1}
\end{figure}

\begin{figure}
  \centering
  \input{figures/ttyping2}
  \input{figures/ttyping3}
  \caption{Effect system of \LC{} (2): well-formedness rules}
  \label{fig:ttyping23}
\end{figure}

\begin{figure}
  \centering
  \input{figures/ttyping4}
  \caption{Effect system of \LC{} (3): typing rules}
  \label{fig:ttyping4}
\end{figure}

The effect system for \LC{} consists of the subtyping relation \(\tau_{{\mathrm{1}}}  \mathrel{<\!:}  \tau_{{\mathrm{2}}}\);
the well-formedness relations \(\Xi \, \ottkw{ok}\), \( \Xi \Vdash \Phi \), \( \Xi \Vdash \varphi \), and
\( \Xi \Vdash \tau \); and typing relation \(\Xi  \vdash  \ottnt{M}  \ottsym{:}  \tau  \mathrel{\mathampersand}  \varphi\).

Well-formedness rules just check a closedness of types by \ruleref{TW-NVar} and
\ruleref{TW-Forall}.

Typing rules are the heart of the effect system.  The judgment
\(\Xi  \vdash  \ottnt{M}  \ottsym{:}  \tau  \mathrel{\mathampersand}  \varphi\) denotes a usual typing relation by the left side of \&
and what names will occur during evaluation of \(\ottnt{M}\) by the right side of
\&.  So, ignoring the right side of \&, rules are already familiar.  For the
effect part, the rules become complicated because of showing meta-properties.

\ruleref{TT-Var}, \ruleref{TT-Abs}, \ruleref{TT-SAbs}, and \ruleref{TT-Fix} are
rather easy.  Corresponding expressions for those are never evaluated (fix-point
operator is evaluated but soon reaches a value); and therefore, those produce no
names examined.  This is why the effect part of those rules becomes empty.  Note
that functions' body is evaluated when an argument is given; so the effect of
the body is recorded on types and added at an application point,
c.f. \ruleref{TT-Abs} and \ruleref{TT-App}.

\ruleref{TT-App}, \ruleref{TT-SApp}, and \ruleref{TT-Choice} have complicated
side-conditions.  To explain the rules, we will starts from the following rather
ideal and simple rule for coordinate choices.
\begin{prooftree}
  \AxiomC{\(\Xi  \vdash  \ottnt{M_{{\mathrm{1}}}}  \ottsym{:}  \tau  \mathrel{\mathampersand}  \varphi_{{\mathrm{1}}}\)}%
  \AxiomC{\(\Xi  \vdash  \ottnt{M_{{\mathrm{2}}}}  \ottsym{:}  \tau  \mathrel{\mathampersand}  \varphi_{{\mathrm{2}}}\)}%
  \AxiomC{\( \Xi \Vdash \Phi \)}%
  \TrinaryInfC{\(\Xi  \vdash   \ottsym{(} \ottnt{M_{{\mathrm{1}}}} \parallel ^{ \Phi } \ottnt{M_{{\mathrm{2}}}} \ottsym{)}   \ottsym{:}  \tau  \mathrel{\mathampersand}  \varphi_{{\mathrm{1}}}  \ottsym{+}  \varphi_{{\mathrm{2}}}  \ottsym{+}  \Phi\)}
\end{prooftree}
This rule just estimate the names which occur along an evaluation and impose no
restriction.  \(\ottnt{M_{{\mathrm{1}}}}\) will cause names represented by the regular expression
\(\varphi_{{\mathrm{1}}}\), \(\ottnt{M_{{\mathrm{2}}}}\) will cause names represented by the regular expression
\(\varphi_{{\mathrm{2}}}\), and the choice itself already causes the name \(\Phi\).  So the
whole effect is the alternation of those.

The next step is to introduce a restriction to the rule.  The idea of
restriction is so simple---coordination never happens if names of
sub-expressions do not overlap.  So the rule will become the following.
\begin{prooftree}
  \AxiomC{\(\Xi  \vdash  \ottnt{M_{{\mathrm{1}}}}  \ottsym{:}  \tau  \mathrel{\mathampersand}  \varphi_{{\mathrm{1}}}\)}%
  \AxiomC{\(\Xi  \vdash  \ottnt{M_{{\mathrm{2}}}}  \ottsym{:}  \tau  \mathrel{\mathampersand}  \varphi_{{\mathrm{2}}}\)}%
  \AxiomC{\( \Xi \Vdash \Phi \)}%
  \AxiomC{\((   \BbbL ( \varphi_{{\mathrm{1}}} )   \uplus   \BbbL ( \varphi_{{\mathrm{2}}} )    \uplus   \BbbL ( \Phi )  )\)}%
  \QuaternaryInfC{\(\Xi  \vdash   \ottsym{(} \ottnt{M_{{\mathrm{1}}}} \parallel ^{ \Phi } \ottnt{M_{{\mathrm{2}}}} \ottsym{)}   \ottsym{:}  \tau  \mathrel{\mathampersand}  \varphi_{{\mathrm{1}}}  \ottsym{+}  \varphi_{{\mathrm{2}}}  \ottsym{+}  \Phi\)}
\end{prooftree}

The last step comes from a technical reason.  The naive side-condition breaks
subject reduction lemma.  We need the following property for the subject
reduction lemma.

\begin{proposition}
  If \(\Xi  \ottsym{,}  \alpha  \vdash  \ottnt{M}  \ottsym{:}  \tau  \mathrel{\mathampersand}  \varphi\) and \(  \ottkw{fnv} ( \Phi )   \subseteq   \ottkw{ndom} ( \Xi )  \), then
  \(\Xi  \vdash   \ottnt{M}  [  \alpha  \coloneq  \Phi  ]    \ottsym{:}   \tau  [  \alpha  \coloneq  \Phi  ]    \mathrel{\mathampersand}   \varphi  [  \alpha  \coloneq  \Phi  ]  \).
\end{proposition}

However, the name substitution for the effect breaks the disjunctivity.  (For
instance, \( \BbbL ( \alpha ) \) and \( \BbbL ( \mdwhtcircle ) \) are disjunctive, but
\( \BbbL (  \alpha  [  \alpha  \coloneq  \mdwhtcircle  ]   ) \) and \( \BbbL (  \mdwhtcircle  [  \alpha  \coloneq  \mdwhtcircle  ]   ) \) are not.)

To amend the problem we adopt more specific side-condition as follows and make
the effect in the conclusion as \( \Phi'   \ottsym{(}  \bar\varphi_{{\mathrm{1}}}  \ottsym{+}  \bar\varphi_{{\mathrm{2}}}  \ottsym{+}  \bar\varphi_{{\mathrm{3}}}  \ottsym{)} \).
\begin{gather*}
    \BbbL ( \varphi_{{\mathrm{1}}} )   \subseteq   \BbbL (  \Phi'   \bar\varphi_{{\mathrm{1}}}  )  \\
    \BbbL ( \varphi_{{\mathrm{2}}} )   \subseteq   \BbbL (  \Phi'   \bar\varphi_{{\mathrm{2}}}  )  \\
    \BbbL ( \Phi )   \subseteq   \BbbL (  \Phi'   \bar\varphi_{{\mathrm{3}}}  )  \\
     \BbbL ( \bar\varphi_{{\mathrm{1}}} )   \uplus   \BbbL ( \bar\varphi_{{\mathrm{2}}} )    \uplus   \BbbL ( \bar\varphi_{{\mathrm{3}}} )  
\end{gather*}
The point is that string variables are gathered into the prefix and name
disjunctivity is guaranteed by the closed effects, which are not altered by a
substitution.  Indeed, this side-condition quite respects the compilation rules,
which create a fresh name by appending the constants \( \mdwhtcircle ,  \mdblkcircle \) to the
tail of seed and a string variable is pushed to the head of the seed.

The \ruleref{TT-Choice} is almost obtained.  We can drive away the first two
subset relation into \ruleref{TT-Sub}.  Additionally, it can be seen that the
effects from left and right side of a choice, namely \(\varphi_{{\mathrm{1}}}\) and \(\varphi_{{\mathrm{2}}}\),
need not be distinct since both sides never collaborate.  So, we can take one
large effect \(\bar\varphi\) which subsume \(\varphi_{{\mathrm{1}}}\) and \(\varphi_{{\mathrm{2}}}\) instead of
\(\bar\varphi_{{\mathrm{1}}}\) and \(\bar\varphi_{{\mathrm{2}}}\), i.e., \(  \BbbL ( \varphi_{{\mathrm{1}}} )   \subseteq   \BbbL (  \Phi'   \bar\varphi  )  \) and
\(  \BbbL ( \varphi_{{\mathrm{2}}} )   \subseteq   \BbbL (  \Phi'   \bar\varphi  )  \).

\ruleref{TT-App} and \ruleref{TT-SApp} are constructed in a similar manner.


\section{Property}\label{sec:cochoice/property}
In the proof of this section, we implicitly use well-known properties about
regular expressions, e.g., \(  \BbbL (   \epsilon    \varphi  )  \ottsym{=}  \BbbL ( \varphi )  \),
\(  \BbbL (  \varphi_{{\mathrm{1}}}   \ottsym{(}  \varphi_{{\mathrm{2}}}  \ottsym{+}  \varphi_{{\mathrm{3}}}  \ottsym{)}  )  \ottsym{=}  \BbbL (  \varphi_{{\mathrm{1}}}    \varphi_{{\mathrm{2}}}  \ottsym{+}  \varphi_{{\mathrm{1}}}   \varphi_{{\mathrm{3}}}   )  \), etc.  This is one reason we
have not fully mechanized the proofs yet.

\subsection{Type Soundness of \LC{}}

We start from investigation for the effect system: the effect system prevents
coordination by \propref{cochoice/non-coordination} and the property is
preserved by the reduction by \propref{cochoice/subject}.

\begin{lemma}[Substitution]
  If \( \Xi ,  \ottmv{x} \mathord: \tau'   \vdash  \ottnt{M}  \ottsym{:}  \tau  \mathrel{\mathampersand}  \varphi\) and \(\Xi  \vdash  \ottnt{M'}  \ottsym{:}  \tau'  \mathrel{\mathampersand}   \msansA \), then \(\Xi  \vdash   \ottnt{M}  [  \ottmv{x}  \coloneq  \ottnt{M'}  ]    \ottsym{:}  \tau  \mathrel{\mathampersand}  \varphi\).
\end{lemma}

\begin{lemma}[Name substitution]\label{prop:cochoice/nsubst}
  If \(\Xi  \ottsym{,}  \alpha  \vdash  \ottnt{M}  \ottsym{:}  \tau  \mathrel{\mathampersand}  \varphi\) and \(  \ottkw{fnv} ( \Phi )   \subseteq   \ottkw{ndom} ( \Xi )  \), then \(\Xi  \vdash   \ottnt{M}  [  \alpha  \coloneq  \Phi  ]    \ottsym{:}   \tau  [  \alpha  \coloneq  \Phi  ]    \mathrel{\mathampersand}   \varphi  [  \alpha  \coloneq  \Phi  ]  \).
\end{lemma}

\begin{lemma}[Subject reduction]\label{prop:cochoice/subject}
  If \(\varnothing  \vdash  \ottnt{M}  \ottsym{:}  \tau  \mathrel{\mathampersand}  \varphi\) and \( \ottnt{M}  \mathrel{\overset{ \Delta }{ \longrightarrow } }  \ottnt{M'} \), then \(\varnothing  \vdash  \ottnt{M'}  \ottsym{:}  \tau  \mathrel{\mathampersand}  \varphi\).
\end{lemma}

\begin{definition}
  Non-coordinated reduction relation, denoted as \(\ottnt{M}  \Longrightarrow  \ottnt{M'}\), is derived
  from the rules in \figref{cochoice/tsemantics} by replacing
  \(\overset{\Delta}{ \longrightarrow }\) with \( \Longrightarrow \) and removing
  \ruleref{TR-WorldL} and \ruleref{TR-WorldR}.
\end{definition}

\begin{lemma}
  If \(\varnothing  \vdash  \ottnt{M}  \ottsym{:}  \tau  \mathrel{\mathampersand}  \varphi\), \( \ottnt{M}  \mathrel{\overset{ \Delta }{ \longrightarrow } }  \ottnt{M'} \), and \( \BbbL ( \varphi )  \uplus \{\omega
  \mid \omega_{+} \in \Delta \vee \omega_{-} \in \Delta\}\); then \(\ottnt{M}  \Longrightarrow  \ottnt{M'}\).
\end{lemma}

\begin{corollary}[Non-coordination]\label{prop:cochoice/non-coordination}
  If \(\varnothing  \vdash  \ottnt{M}  \ottsym{:}  \tau  \mathrel{\mathampersand}  \varphi\) and \( \ottnt{M}  \mathrel{\overset{ \varnothing }{ \longrightarrow } }  \ottnt{M'} \), then \(\ottnt{M}  \Longrightarrow  \ottnt{M'}\).
\end{corollary}

\subsection{Soundness of Compilation}

Towards the goal we show a well-typed expression in \LP{} is mapped into a
well-typed expression in \LC{} by the compilation.

\begin{definition}[Compilation for types]
  Compilation for types is given as follows.
  \begin{align*}
      \lBrack \ottkw{nat} \rBrack   &=  \ottkw{nat} \\
      \lBrack T_{{\mathrm{1}}}  \rightarrow  T_{{\mathrm{2}}} \rBrack   &=    \lBrack T_{{\mathrm{1}}} \rBrack  \overset{  \msansA  }{ \rightarrow }  \forall \alpha .^{   \alpha   \ottsym{(}   \mdwhtcircle \mdblkcircle   \ottsym{)}  ^\ast  }  \lBrack T_{{\mathrm{2}}} \rBrack    \\
  \end{align*}
\end{definition}

\begin{definition}[Compilation for typing environment]
  Compilation for typing environment is given by the following obvious way.
  \begin{align*}
     \lBrack \varnothing \rBrack  &=  \varnothing \\
     \lBrack  \Gamma , \ottmv{x} \mathord: T  \rBrack  &=   \lBrack \Gamma \rBrack  ,  \ottmv{x} \mathord:  \lBrack T \rBrack  
  \end{align*}
\end{definition}

\begin{lemma}
  \( \varnothing \Vdash  \lBrack T \rBrack  \).
\end{lemma}

\begin{lemma}
  \( \lBrack \Gamma \rBrack  \, \ottkw{ok}\).
\end{lemma}

\begin{lemma}\label{prop:cochoice/trans-sound}
  If \(\Gamma  \vdash  \ottnt{e}  \ottsym{:}  T\), then \( \lBrack \Gamma \rBrack   \ottsym{,}  \alpha  \vdash   \lBrack \ottnt{e} \rBrack^{ \alpha }_{ \bar\Phi }   \ottsym{:}   \lBrack T \rBrack   \mathrel{\mathampersand}     \alpha   \bar\Phi    \ottsym{(}   \mdwhtcircle \mdblkcircle   \ottsym{)}  ^\ast \).
  \begin{proof}
    The proof is by induction on the given derivation.
  \end{proof}
\end{lemma}

\subsection{Bisimulation}

Next we show a well-typed expression in \LC{} behaves as same as the expression
of \LP{} which is obtained by erasing the name related parts.

\begin{definition}[strong bisimulation between \LP{} and \LC{}]
  A binary relation \(R\) between \LP{} and \LC{} expressions is called \emph{strong
    bisimulation} iff the following conditions hold.
  \begin{itemize}
  \item If \(\ottnt{e} \mathrel{R} \ottnt{M}\) and \(\ottnt{e}  \longrightarrow  \ottnt{e'}\), then \( \ottnt{M}  \mathrel{\overset{ \varnothing }{ \longrightarrow } }  \ottnt{M'} \) and
    \(\ottnt{e'} \mathrel{R} \ottnt{M'}\).
  \item If \(\ottnt{e} \mathrel{R} \ottnt{M}\) and \( \ottnt{M}  \mathrel{\overset{ \varnothing }{ \longrightarrow } }  \ottnt{M'} \), then \(\ottnt{e}  \longrightarrow  \ottnt{e'}\) and
    \(\ottnt{e'} \mathrel{R} \ottnt{M'}\).
  \end{itemize}
\end{definition}

\begin{definition}[weak bisimulation between \LP{} and \LC{}]
  A binary relation \(R\) between \LP{} and \LC{} expressions is called \emph{weak
    bisimulation} iff the following conditions hold.
  \begin{itemize}
  \item If \(\ottnt{e} \mathrel{R} \ottnt{M}\) and \(\ottnt{e}  \longrightarrow  \ottnt{e'}\), then \( \ottnt{M}  \mathrel{\overset{ \varnothing }{\longrightarrow}{}^\ast}  \ottnt{M'} \) and
    \(\ottnt{e'} \mathrel{R} \ottnt{M'}\).
  \item If \(\ottnt{e} \mathrel{R} \ottnt{M}\) and \( \ottnt{M}  \mathrel{\overset{ \varnothing }{ \longrightarrow } }  \ottnt{M'} \), then \(\ottnt{e}  \longrightarrow^\ast  \ottnt{e'}\) and
    \(\ottnt{e'} \mathrel{R} \ottnt{M'}\).
  \end{itemize}
\end{definition}

\begin{definition}[weak bisimulation for \LP{}]
  We also call a binary relation \(R\) between \LP{} expressions weak
  bisimulation iff the following conditions hold.
  \begin{itemize}
  \item If \(\ottnt{e_{{\mathrm{1}}}} \mathrel{R} \ottnt{e_{{\mathrm{2}}}}\) and \(\ottnt{e_{{\mathrm{1}}}}  \longrightarrow  \ottnt{e'_{{\mathrm{1}}}}\), then \(\ottnt{e_{{\mathrm{2}}}}  \longrightarrow^\ast  \ottnt{e'_{{\mathrm{2}}}}\) and \(\ottnt{e'_{{\mathrm{1}}}} \mathrel{R} \ottnt{e'_{{\mathrm{2}}}}\).
  \item If \(\ottnt{e_{{\mathrm{1}}}} \mathrel{R} \ottnt{e_{{\mathrm{2}}}}\) and \(\ottnt{e_{{\mathrm{2}}}}  \longrightarrow  \ottnt{e'_{{\mathrm{2}}}}\), then \(\ottnt{e_{{\mathrm{1}}}}  \longrightarrow^\ast  \ottnt{e'_{{\mathrm{1}}}}\) and \(\ottnt{e'_{{\mathrm{1}}}} \mathrel{R} \ottnt{e'_{{\mathrm{2}}}}\).
  \end{itemize}
\end{definition}

\begin{figure}
  \centering
  \begin{align*}
      \lfloor \ottmv{x} \rfloor   &=  \ottmv{x} \\
      \lfloor \ottnt{M_{{\mathrm{1}}}} \, \ottnt{M_{{\mathrm{2}}}} \rfloor   &=   \lfloor \ottnt{M_{{\mathrm{1}}}} \rfloor  \,  \lfloor \ottnt{M_{{\mathrm{2}}}} \rfloor  \\
      \lfloor \lambda  \ottmv{x}  \ottsym{.}  \ottnt{M} \rfloor   &=  \lambda  \ottmv{x}  \ottsym{.}   \lfloor \ottnt{M} \rfloor  \\
      \lfloor \ottnt{M} \, \Phi \rfloor   &=   \lfloor \ottnt{M} \rfloor  \, \lambda  \ottmv{x}  \ottsym{.}  \ottmv{x} \\
      \lfloor  \Lambda  \alpha . \ottnt{M}  \rfloor   &=  \lambda  \ottmv{x}  \ottsym{.}   \lfloor \ottnt{M} \rfloor   & \text{where \(\ottmv{x}\) is fresh}\\
      \lfloor  \mu \ottmv{f} . \ottnt{M}  \rfloor   &=   \mu \ottmv{f} .  \lfloor \ottnt{M} \rfloor   \\
      \lfloor  \ottsym{(} \ottnt{M_{{\mathrm{1}}}} \parallel ^{ \Phi } \ottnt{M_{{\mathrm{2}}}} \ottsym{)}  \rfloor   &=  \ottsym{(}   \lfloor \ottnt{M_{{\mathrm{1}}}} \rfloor   \parallel   \lfloor \ottnt{M_{{\mathrm{2}}}} \rfloor   \ottsym{)} 
  \end{align*}
  \caption{Name erasure function}
  \label{fig:erase}
\end{figure}

\begin{definition}
  We define the \emph{name erasure} function \(\lfloor\cdot\rfloor\) from \LC{}
  expressions into \LP{} expressions as in \figref{erase}.  The important point
  of the definition is that name abstractions and applications are replaced by
  dummy lambda abstractions and applications.  If we do not do that, i.e., just
  erase the name abstractions and applications, it will happens that a value of
  \LC{}, which cannot be evaluated, is evaluated in \LP{} after applying the
  name erasure function.  (Consider \( \Lambda  \alpha . \ottsym{(}  \lambda  \ottmv{x}  \ottsym{.}  \ottmv{x}  \ottsym{)} \, \lambda  \ottmv{x}  \ottsym{.}  \ottmv{x} \) and name erased
  expression \(\ottsym{(}  \lambda  \ottmv{x}  \ottsym{.}  \ottmv{x}  \ottsym{)} \, \lambda  \ottmv{x}  \ottsym{.}  \ottmv{x}\).)
\end{definition}

\begin{lemma}
  \(  \lfloor  \ottnt{M}  [  \ottmv{x}  \coloneq  \ottnt{M'}  ]   \rfloor  \ottsym{=}   \lfloor \ottnt{M} \rfloor   [  \ottmv{x}  \coloneq   \lfloor \ottnt{M'} \rfloor   ]  \).
  \begin{proof}
    The proof is routine by structural induction on \(\ottnt{M}\).
  \end{proof}
\end{lemma}

\begin{lemma}
  \(  \lfloor  \ottnt{M}  [  \alpha  \coloneq  \Phi  ]   \rfloor  \ottsym{=}  \lfloor \ottnt{M} \rfloor  \).
  \begin{proof}
    The proof is routine by structural induction on \(\ottnt{M}\).
  \end{proof}
\end{lemma}

\begin{lemma}\label{prop:cochoice/red-pred}
  If \(\varnothing  \vdash  \ottnt{M}  \ottsym{:}  \tau  \mathrel{\mathampersand}  \varphi\) and \( \lfloor \ottnt{M} \rfloor   \longrightarrow  \ottnt{e'}\), then \( \ottnt{M}  \mathrel{\overset{ \varnothing }{ \longrightarrow } }  \ottnt{M'} \)
  and \(  \lfloor \ottnt{M'} \rfloor  \ottsym{=} \ottnt{e'} \).
  \begin{proof}
    The proof is by induction on the given derivation of \(\varnothing  \vdash  \ottnt{M}  \ottsym{:}  \tau  \mathrel{\mathampersand}  \varphi\).  Note that well-typedness of \(\ottnt{M}\) is necessary because function
    applications and name applications are collapsed by name erasing.
  \end{proof}
\end{lemma}

\begin{lemma}\label{prop:cochoice/pred-red}
  If \(\ottnt{M}  \Longrightarrow  \ottnt{M'}\), then \( \lfloor \ottnt{M} \rfloor   \longrightarrow   \lfloor \ottnt{M'} \rfloor \).
  \begin{proof}
    The proof is by induction on the given derivation.
  \end{proof}
\end{lemma}

\begin{corollary}\label{prop:cochoice/cred-red}
  If \(\varnothing  \vdash  \ottnt{M}  \ottsym{:}  \tau  \mathrel{\mathampersand}  \varphi\) and \( \ottnt{M}  \mathrel{\overset{ \varnothing }{ \longrightarrow } }  \ottnt{M'} \), then \( \lfloor \ottnt{M} \rfloor   \longrightarrow   \lfloor \ottnt{M'} \rfloor \).
  \begin{proof}
    This is a corollary of \propref{cochoice/non-coordination} and \propref{cochoice/pred-red}.
  \end{proof}
\end{corollary}

\begin{definition}
  We give the binary relation between source expressions and target expressions,
  written \(\ottnt{e}  \sim  \ottnt{M}\) as follows.
  \begin{gather*}
    \ottnt{e}  \sim  \ottnt{M} \iff \varnothing  \vdash  \ottnt{M}  \ottsym{:}  \tau  \mathrel{\mathampersand}  \varphi \wedge   \lfloor \ottnt{M} \rfloor  \ottsym{=} \ottnt{e} 
  \end{gather*}
\end{definition}

\begin{corollary}
  \( \sim \) is a strong bisimulation.
  \begin{proof}
    This is a corollary of \propref{cochoice/subject}, \propref{cochoice/red-pred}, and \propref{cochoice/cred-red}.
  \end{proof}
\end{corollary}

\begin{corollary}
  If \(\varnothing  \vdash  \ottnt{e}  \ottsym{:}  T\), then \( \lfloor   \lBrack \ottnt{e} \rBrack^{ \alpha }_{ \bar\Phi }   [  \alpha  \coloneq   \epsilon   ]   \rfloor   \sim    \lBrack \ottnt{e} \rBrack^{ \alpha }_{ \bar\Phi }   [  \alpha  \coloneq   \epsilon   ]  \).
  \begin{proof}
    This is a corollary of \propref{cochoice/nsubst} and \propref{cochoice/trans-sound}.
  \end{proof}
\end{corollary}

\begin{figure}
  \centering
  \begin{align*}
      \lceil \ottmv{x} \rceil   &=  \ottmv{x} \\
      \lceil \ottnt{e_{{\mathrm{1}}}} \, \ottnt{e_{{\mathrm{2}}}} \rceil   &=   \lceil \ottnt{e_{{\mathrm{1}}}} \rceil  \,  \lceil \ottnt{e_{{\mathrm{2}}}} \rceil  \, \lambda  \ottmv{x}  \ottsym{.}  \ottmv{x} \\
      \lceil \lambda  \ottmv{x}  \ottsym{.}  \ottnt{e} \rceil   &=  \lambda  \ottmv{x}  \ottsym{.}  \lambda  \ottmv{y}  \ottsym{.}   \lceil \ottnt{e} \rceil   & \text{where \(\ottmv{y}\) is fresh}\\
      \lceil  \mu \ottmv{f} . \ottnt{e}  \rceil   &=   \mu \ottmv{f} .  \lceil \ottnt{e} \rceil   \\
      \lceil \ottsym{(}  \ottnt{e_{{\mathrm{1}}}}  \parallel  \ottnt{e_{{\mathrm{2}}}}  \ottsym{)} \rceil   &=  \ottsym{(}   \lceil \ottnt{e_{{\mathrm{1}}}} \rceil   \parallel   \lceil \ottnt{e_{{\mathrm{2}}}} \rceil   \ottsym{)} 
  \end{align*}
  \caption{Pseudo compilation}
  \label{fig:skel}
\end{figure}

\begin{definition}
  We define the \emph{pseudo compilation} from/to \LP{} expressions as shown in
  \figref{skel}.  This function is not essential for our discussion, but we use
  the function for convenience writing in the following.
\end{definition}

\begin{lemma}
  \(  \lfloor  \lBrack \ottnt{e} \rBrack^{ \alpha }_{ \bar\Phi }  \rfloor  \ottsym{=}  \lceil \ottnt{e} \rceil  \).
\end{lemma}

\begin{lemma}
  \(  \lceil  \ottnt{e}  [  \ottmv{x}  \coloneq  \ottnt{e'}  ]  \rceil  \ottsym{=}   \lceil \ottnt{e} \rceil   [  \ottmv{x}  \coloneq   \lceil \ottnt{e'} \rceil   ]  \).
\end{lemma}

\begin{lemma}
  If \( \lceil \ottnt{e} \rceil   \longrightarrow  \ottnt{e'}\), then \(\ottnt{e}  \longrightarrow^\ast  \ottnt{e''}\) and
  \(  \lceil \ottnt{e''} \rceil  \ottsym{=} \ottnt{e'} \).
\end{lemma}

\begin{lemma}
  If \(\ottnt{e}  \longrightarrow  \ottnt{e'}\), then \( \lceil \ottnt{e} \rceil   \longrightarrow^\ast  \ottnt{e''}\) and
  \(  \lceil \ottnt{e'} \rceil  \ottsym{=} \ottnt{e''} \).
\end{lemma}

\begin{definition}
  We give the binary relation between \LP{} expressions, denoted by \(\approx\),
  as \(\ottnt{e}  \approx   \lceil \ottnt{e} \rceil \).
\end{definition}

\begin{corollary}
  The binary relation \(\ottnt{e}  \approx   \lceil \ottnt{e} \rceil \) is a weak bisimulation.
\end{corollary}

\begin{theorem}
  If \(\varnothing  \vdash  \ottnt{e}  \ottsym{:}  T\), then \(\ottnt{e} \approx \vysmwhtcircle \sim   \lBrack \ottnt{e} \rBrack^{ \alpha }_{  \epsilon  }   [  \alpha  \coloneq   \epsilon   ]  \).
\end{theorem}

The result is a bit blurred since expressions before and after compilation are
related by two relations.  More directly, the correspondence can be shown as follows.

\begin{gather*}
  \begin{aligned}
    \ottnt{e} & \longrightarrow^\ast  \ottnt{e'}\\
      \lBrack \ottnt{e} \rBrack^{ \alpha }_{  \epsilon  }   [  \alpha  \coloneq   \epsilon   ]   &\mathrel{\overset{ \varnothing }{ \longrightarrow^\ast }} \ottnt{M'}\\
  \end{aligned}
  \quad\text{and}\quad
    \lceil \ottnt{e'} \rceil  \ottsym{=}  \lfloor \ottnt{M'} \rfloor  
\end{gather*}


\section{Conclusion}
We give a compilation algorithm from \LP{}, a simply typed
lambda calculus with nondeterministic choices, into \LC{}, a simply typed lambda
calculus with coordinated choices; and show the compilation is sound and
correct.
%

\bibliographystyle{plain}
\bibliography{reference}


\end{document}